\documentclass{ptapap}

\def\hd{HD\,201433}
\def\d{\,d$^{-1}$}
\def\sun{\hbox{$_\odot$}}

\author{Thomas Kallinger}[IFA]
\author{Werner W. Weiss}[IFA]
\author{BEST}[BEST]
\affil[IFA]{Institute for Astrophysics (IfA), University of Vienna, T\"urkenschanzstrasse 17, 1180 Vienna, Austria}
\affil[BEST]{Bright Target Explorer (BRITE) Executive Science Team}

\title{Bayesian frequency analysis of HD\,201433 observations with BRITE}
\begin{document}

\maketitle

\begin{abstract}
Multiple oscillation frequencies separated by close to or less than the formal frequency resolution of a data set are a serious problem in the frequency analysis of time series data. We present a new and fully automated Bayesian approach that searches for close frequencies in time series data and assesses their significance by comparison to no signal and a mono-periodic signal. We extensively test the approach with synthetic data sets and apply it to the 156 days-long high-precision BRITE photometry of the SPB star \hd , for which we find a sequence of nine statistically significant rotationally split dipole modes.
\end{abstract}

\section{Introduction}
Even in the days of high-precision photometry from space-based instruments like \textit{Kepler} \citep{bor10} and BRITE-Constellation \citep{weiss2014} the innocently-looking \textit{frequency analysis} of time-series data can be quite difficult. This is especially the case in the presence of multiple frequencies that are separated by close to or less than the formal frequency (Rayleigh-) resolution of the time series ($1/T$, with $T$ being the data set length) like rotationally split modes of slowly rotating stars or the densely packed gravity mode spectrum of massive stars. 

Classical frequency analysis (with, e.g., \textit{SigSpec} - \citealt{reegen2007}, or \textit{Period04} - \citealt{lenz2005}) is often based on a strict pre-whitening sequence (i.e., compute Fourier amplitude spectrum $\rightarrow$ least-squares fit around the highest-peak frequency $\rightarrow$ subtract the fit from the data $\rightarrow$ start from the beginning with the residuals). A problem in such an approach is the presence of close-frequency pairs (or multiplets) because assuming a mono-periodic signal in the vicinity of the considered Fourier peak yields a frequency that corresponds to the weighted average of the intrinsic frequency multiplet and pre-whitening this ``wrong'' signal causes artificial peaks in the spectrum. Furthermore is it difficult (or often impossible) to objectively rate the significance of the result. User experience is therefore a critical parameter in the analysis, which is an unsatisfactory situation.

As usual, a probabilistic approach helps to tackle such problems. We developed a fully automated Bayesian algorithm that searches for close frequencies in  time series data and tests their statistical significance by comparison to a fit with constant (i.e., no periodic) signal and a fit with a mono-periodic signal. We extensively tested the algorithm with synthetic data and finally applied it to the BRITE observations of the slowly pulsating B (SPB) star \hd , in which we find a sequence of nine pairs of close frequencies that are statistically significant.

\section{Bayesian frequency analysis} \label{sec:freqAna}

%SPB stars are expected to show long-period g modes in a frequency range of up to a few cycles per day. This is well separated from residual instrumental signal and alias peaks due to the orbital frequency of BTr of $\sim$14.7\d (and multiples of it). We compute the Fourier amplitude spectrum of the unbinned light curve and find no significant peak between 2\d and the Nyquist frequency which can not be associated to the satellites orbital frequency. The pulsation spectrum (see Fig.\,\ref{fig:fspec}) of \hd\ shows the strongest peaks between about 0.5 -- 1\d, and another group of peaks between about 1.5 -- 2\d. Above that no significant power can be found.

To extract the frequencies of pulsation modes (and any other possible light variations) we performed a pre-whitening procedure to iteratively decompose the time series into its harmonic components. The procedure performs the following steps:
\begin{itemize}
	\item Compute the Fourier amplitude spectrum for an user-defined frequency range and determine the frequency with the highest amplitude.
	\item Fit $N$ functions, $F_{(t,n)} = \sum_{i=1}^n A_i \sin{[2\pi(f_it+\Phi_i)]} + c$, to the time series, where $n$ incrementally increases from 1 to $N$ so that, in total, $N$ models with 1, 2, ..., $N$ sinusoidal components\footnote{$N$ is a user-defined parameter but usually a value of 3 is sufficient to reproduce the data.} are fit to the data. $A$, $f$, and $\Phi$ are the amplitude, frequency, and phase of the $i$th component, respectively. The parameter $c$ serves as an offset to ensure that $\int_{T}{F_{(t)}dt=0}$ even if the duration $T$ of the time series is not an integer multiple of the signal period. For the fit we use a Bayesian nested sampling algorithm \citep[\textsc{MultiNest;}][]{feroz2009}, and allow the individual frequencies to vary around the initial frequency by $\pm 2/T$, and the amplitudes between 0 and twice the initial amplitude from the amplitude spectrum. The phases have no initial constraints and can vary between 0 and 1.
	\item To rate if a signal is statistically significant (i.e., not be due to noise) and if so, which model represents the data best, we compute the model probability ($p_n$) by comparing the global evidences ($z_n$) of the fits (as delivered by \textsc{MultiNest}) to those of a fit with a constant factor ($z_c$). If $p = \sum z_n/(z_c + \sum z_n) > 0.95$ we consider the solution as real\footnote{In probability theory already an odds ratio of 10:1 (i.e., p=0.9) is considered as strong evidence \citep{jeffreys98}} and not to be due to noise. If so, the best-fit model is then the model with $p_n = z_n / \sum z_n > 0.95$. This means that in order to be accepted, a multiperiodic solution needs to considerably better fit the data than the monochromatic solution. Our approach for the statistical significance of a signal compares well to classical approaches like a SNR $>$ 4 \citep[e.g.,][]{breger1993} but has the advantage of providing an actual statistical statement that is based on the data only and that allows us to discriminate between mono- and multi-periodic solutions for closely separated frequencies.
	\item The best-fit parameters and their 1$\sigma$ uncertainties are then computed from the marginalized posterior distribution functions as delivered by \textsc{MultiNest}.
	\item The best-fit model is subtracted from the time series and the procedure starts from the beginning.
\end{itemize}

The procedure stops when $p$ drops below 0.66 (corresponding to a weak evidence) but we accept only those frequencies with $p>0.95$. We note that the frequency, amplitude, and phase uncertainties that are computed from the posterior probability distributions compare well with uncertainties determined from other criteria \citep[e.g.,][]{kal2008}.

\subsection{Tests with synthetic data}

We extensively tested our approach with synthetic time series. We investigated in particular realistic limits for the resolution of frequencies separated by less than the formal, classical frequency resolution $1/T$, and as a function of their amplitude. The synthetic data are based on the sampling and noise characteristics of the BRITE observations of \hd\ (see Sec.\,\ref{sec:britephot}) and therefore have a frequency resolution of about 1/156d $\simeq$ 0.0064\d . The synthetic data include a pair of close frequencies randomly separated by 0.05 to 2 times $1/T$. Their amplitude ratio is fixed to 1:2 (which roughly corresponds to what we find for \hd ) with the amplitude of the larger component randomly set between 0 and 10\,ppt. For comparison we add a third (well separated by more than 10/T) frequency with the same amplitude as the low-amplitude component. All phases are random and independent.

\begin{figure}[h]
\centering
\includegraphics[width=0.6\textwidth]{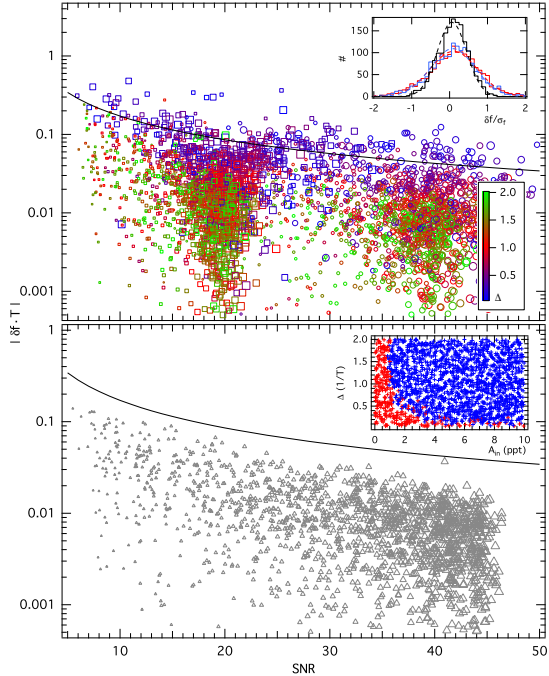}
\caption{Absolute deviation (in units of the frequency resolution) between the input frequencies and the frequencies determined with \textsc{MultiNest} as a functions of their SNR for 2\,000 simulated data sets each including a pair of close frequencies (top panel) and a single frequency (bottom panel). The separation between the pair of close frequencies (in units of the frequency resolution) is colour-coded and the symbol size indicates the input amplitude. Open squares and circles indicate the small and large component of the frequency pair, respectively (with a fixed amplitude ratio of 2:1). Black lines give the upper frequency error limit for a mono-periodic signal as defined by \citet{kal2008}. The insert in the top panel shows histograms (solid lines) of $\delta$f scaled to the \textsc{MultiNest} uncertainties and Gaussian fits to them (dashed lines), with the red and blue lines corresponding to the small and large amplitude components, respectively, and black lines to the single frequency. The insert in the bottom panel shows the input parameters of the simulations with blue symbols indicating that our algorithm has correctly identified a pair of close frequencies and red symbols indicating a misidentification as a single frequency.}
\label{fig:fsim}
\end{figure}

We then apply our frequency analysis algorithm to 2\,000 synthetic data sets in order to see to what extent we can reliably identify pairs of close frequencies  and reconstruct their input parameters. The result is illustrated in Fig.\,\ref{fig:fsim}, where we show the deviation between the input frequency of the simulation and the actually determined frequency ($\delta$f) as a function of their SNR for the close-frequency pair (top panel). For a comparison we do the same for a single frequency (bottom panel). The found frequency deviations ($\delta$f) obviously depend on the SNR but also on how far the two frequencies are separated (colour coded $\Delta$). If the two frequencies are separated by more than about $1.5/T$ the individual frequencies can be reconstructed about as accurate as a ``mono-periodic'' signal, i.e., the two frequencies do not influence each other in the analysis. Only for smaller separations $\delta$f starts to increase but even for very close frequencies ($\Delta$ < 0.5/T) the mismatch between input and output frequency rarely exceeds 0.1/T.

Another question is to which limit we can reliably distinguish between a single frequency and a pair of close frequencies. This is shown in the insert of the bottom panel in Fig.\,\ref{fig:fsim}, where we plot the input parameters of our simulation and colour-code the solution found by our algorithm. If  the input frequencies are separated by about $1/T$ (which is what we are roughly dealing with in the real data of \hd ) we can reliably (>99.9\%) distinguish between a single frequency and a pair of close frequencies down to an input amplitude (of the stronger component) of about 1\,ppt. Only for frequencies closer than about $0.5/T$ the detection limit rapidly increases and below about $0.1/T$ a discrimination is no longer possible.

Finally, we also compare the parameter uncertainties as delivered by \textsc{MultiNest} to the actually found deviations. Histograms of the frequency deviations scaled to the frequency uncertainty are shown in the insert of the top panel in Fig.\,\ref{fig:fsim}. If the frequency uncertainties from \textsc{MultiNest} happen to be real $1\sigma$ uncertainties, Gaussian fits to the histograms would have a width of one. But in fact they show that \textsc{MultiNest} overestimates the frequency uncertainty by about 40\% for the pair of close frequencies and by about 60\% for the single frequency. We assume that this is due to correlations between the fitted parameters. To be on the conservative side we still treat the \textsc{MultiNest} uncertainties as $1\sigma$ uncertainties. 

For sake of completeness we note that we can reconstruct the input amplitudes to within about 8\% and 16\% for the strong and weak component of the close-frequency pair, respectively, and to within about 4\% for the single frequency. The input phases are reconstructed to within about 3\% and 4\% for the strong and weak component of the close-frequency pair, respectively, and to within about 1.6\% for the single frequency. 

\section{The case of \hd} \label{sec:britephot}

A practical application for our new Bayesian frequency analysis tool are the BRITE observations of the SPB star \hd . SPB stars are non-radial multi-periodic oscillators on the main sequence between spectral type B3 and B9, with an effective temperature of about 11,000 -- 22,000 K, and a mass of 2.5 -- 8\,M\sun\ \citep[e.g.,][]{aerts2010}. They oscillate in high-order gravity modes with frequencies typically ranging from 0.5 to 2\d . Consecutive radial order gravity modes of the same spherical degree are expected to be equally spaced in period, and deviations from this regular pattern carry information about physical processes in the near-core region \citep[e.g.,][]{miglio2008}.

\hd\ was one of the targets in the BRITE-Constellation Cygnus\,II field and was observed with BRITE-Toronto\footnote{The Canadian satellite BRITE-Toronto (BTr) was launched on June 19, 2014, into a slightly elliptical and almost Sun-synchronous orbit and is equipped with a \textit{red} filter} for about 156 days in June--October 2015 typically 48 times per 98\,min BRITE orbit with an average cadence of 5\,s exposures every 20.3\,s. Details about the flux extraction and data post-processing may be found in Kallinger et al (2016, submitted).  

\begin{figure}[t]
\centering
\includegraphics[width=0.7\textwidth]{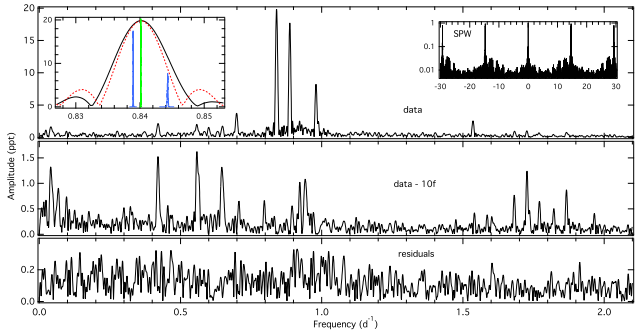}
\caption{Fourier amplitude spectrum of the BTr light curve of \hd\ (top). The middle and bottom panels show the spectrum after prewithening of 10 and all significant frequencies, respectively. The right insert in the top panel shows the spectral window function. The left insert gives the original spectrum (black line) and the spectral window (red dashed line) plotted on top of the main peak. The green and blue peaks indicate the posterior parameter distributions (arbitralily scaled) of a single and multiple sine fit, repsectively.}
\label{fig:fourier}
\end{figure}

In Fig.\,\ref{fig:fourier} we show the amplitude spectrum before, during, and after the pre-whitening of the 29 frequencies that are found to be significant by our algorithm. We find that some peaks in the amplitude spectrum are broader than one could expect from spectral window broadening indicating the presence of more than one (closely separated) frequency. In fact, our frequency analysis algorithm identifies 9 ``features'' in the spectrum that reveal statistically significant closely separated frequencies. An example is illustrated in the left insert of Fig.\,\ref{fig:fourier}, where we show the posterior parameter distributions of the one-frequency and two-frequency model fits for the highest-amplitude peak in the spectrum. The evidence of the latter is orders of magnitude better than for the former model, indicating that more than one frequency is needed to reproduce the data in this frequency range.

\begin{figure}[t]
\centering
\includegraphics[width=\textwidth]{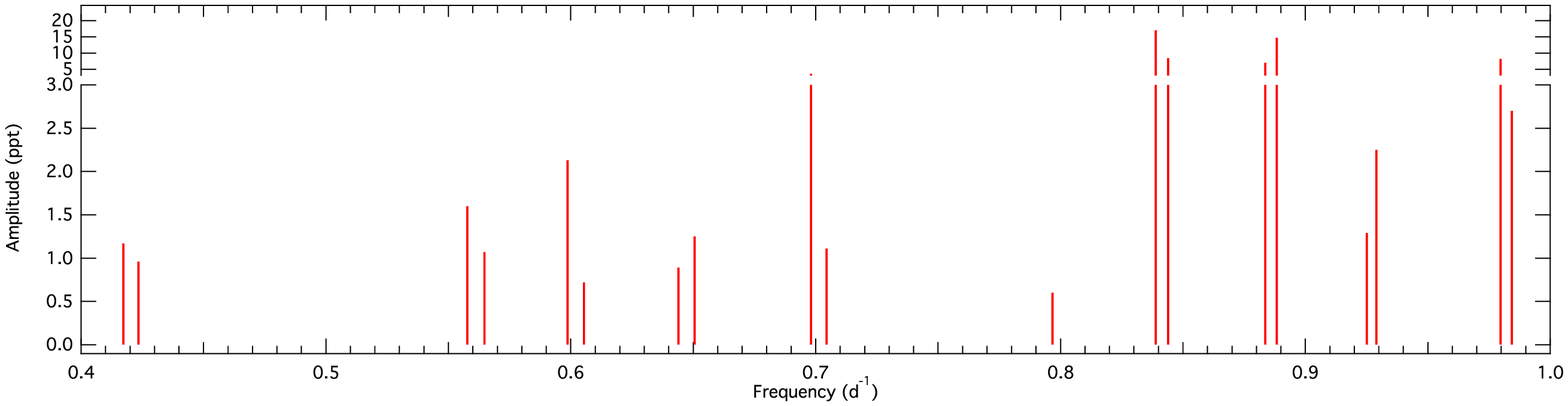}
\caption{Schematic view of the frequency duplets found in \hd .}
\label{fig:freq}
\end{figure}

\section{Conclussions}
We have presented a new Bayesian frequency analysis tool that searches for close frequency multiplets in time series data and rates their statistical significance in a fully automatic way. We have tested the algorithm with synthetic data and applied it to the BTr observations of \hd , for which we find a sequence of nine significant frequency duplets (Fig.\,\ref{fig:freq}). This exciting new BRITE result is consistent with rotationally split dipole modes and indicates a non-rigid internal rotation profile of the star -- a very rare insight into a hot star. For more details and the subsequent asteroseismic analysis including investigations on the internal rotation profile of \hd\ we refer to Kallinger et al. (2016, submitted).

\acknowledgements{The authors are grateful for funding via the Austrian Space Application Programme (ASAP) of the Austrian Research Promotion Agency (FFG) and BMVIT. The paper is based on data collected by the BRITE Constellation satellite mission, designed, built, launched, operated and supported by the Austrian Research Promotion Agency (FFG), the University of Vienna, the Technical University of Graz, the Canadian Space Agency (CSA), the University of Toronto Institute for Aerospace Studies (UTIAS), the Foundation for Polish Science \& Technology (FNiTP MNiSW), and National Science Centre (NCN).
}

\bibliographystyle{ptapap}
\bibliography{HD201433}

\begin{thebibliography}{10}
\providecommand{\natexlab}[1]{#1}
\providecommand{\url}[1]{\texttt{#1}}
\providecommand{\urlprefix}{URL }
\providecommand{\eprint}[2][]{\url{#2}}

\bibitem[{{Aerts} et~al.(2010){Aerts}, {Christensen-Dalsgaard}, \&
  {Kurtz}}]{aerts2010}
{Aerts}, C., {Christensen-Dalsgaard}, J., {Kurtz}, D.~W., {Asteroseismology}
  (2010)

\bibitem[{{Borucki} et~al.(2010)}]{bor10}
{Borucki}, W.~J., et~al., \emph{{Kepler Planet-Detection Mission: Introduction
  and First Results}}, \emph{Science} \textbf{327}, 977 (2010)

\bibitem[{{Breger} et~al.(1993)}]{breger1993}
{Breger}, M., et~al., \emph{{Nonradial Pulsation of the Delta-Scuti Star
  Bu-Cancri in the Praesepe Cluster}}, \emph{\aap} \textbf{271}, 482 (1993)

\bibitem[{{Feroz} et~al.(2009){Feroz}, {Hobson}, \& {Bridges}}]{feroz2009}
{Feroz}, F., {Hobson}, M.~P., {Bridges}, M., \emph{{MULTINEST: an efficient and
  robust Bayesian inference tool for cosmology and particle physics}},
  \emph{\mnras} \textbf{398}, 1601 (2009), \eprint{0809.3437}

\bibitem[{Jeffreys(1998)}]{jeffreys98}
Jeffreys, H., Theory of probability, Oxford Classic Texts in the Physical
  Sciences, xii+459 , The Clarendon Press Oxford University Press, New York
  (1998), reprint of the 1983 edition

\bibitem[{{Kallinger} et~al.(2008){Kallinger}, {Reegen}, \& {Weiss}}]{kal2008}
{Kallinger}, T., {Reegen}, P., {Weiss}, W.~W., \emph{{A heuristic derivation of
  the uncertainty for frequency determination in time series data}},
  \emph{\aap} \textbf{481}, 571 (2008), \eprint{0801.0683}

\bibitem[{{Lenz} \& {Breger}(2005)}]{lenz2005}
{Lenz}, P., {Breger}, M., \emph{{Period04 User Guide}}, \emph{Communications in
  Asteroseismology} \textbf{146}, 53 (2005)

\bibitem[{{Miglio} et~al.(2008){Miglio}, {Montalb{\'a}n}, {Noels}, \&
  {Eggenberger}}]{miglio2008}
{Miglio}, A., {Montalb{\'a}n}, J., {Noels}, A., {Eggenberger}, P.,
  \emph{{Probing the properties of convective cores through g modes: high-order
  g modes in SPB and {$\gamma$} Doradus stars}}, \emph{\mnras} \textbf{386},
  1487 (2008), \eprint{0802.2057}

\bibitem[{{Reegen}(2007)}]{reegen2007}
{Reegen}, P., \emph{{SigSpec. I. Frequency- and phase-resolved significance in
  Fourier space}}, \emph{\aap} \textbf{467}, 1353 (2007),
  \eprint{physics/0703160}

\bibitem[{{Weiss} et~al.(2014)}]{weiss2014}
{Weiss}, W.~W., et~al., \emph{{BRITE-Constellation: Nanosatellites for
  Precision Photometry of Bright Stars}}, \emph{\pasp} \textbf{126}, 573
  (2014), \eprint{1406.3778}

\end{thebibliography}

\end{document}